# Rupture of liquid bridges on porous tips: Competing mechanisms of spontaneous imbibition and stretching

Si Suo[1] and Yixiang Gan[1]

[1] School of Civil Engineering, The University of Sydney, Sydney, NSW 2006, Australia

**Graphical abstract:**

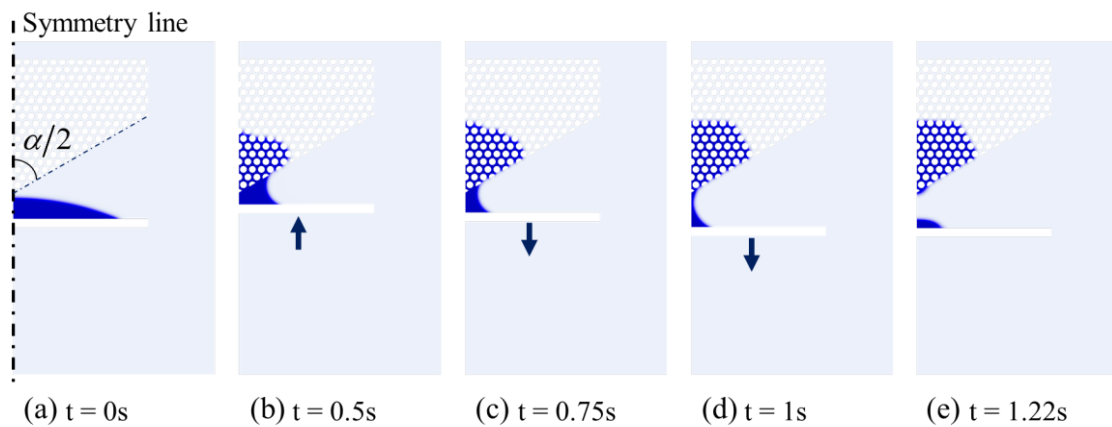

(a) t = 0s    (b) t = 0.5s    (c) t = 0.75s    (d) t = 1s    (e) t = 1.22s

**ABSTRACT:** Liquid bridges are commonly encountered in nature and the liquid transfer induced by their rupture are widely used in various industrial applications. In this work, with the focus on the porous tip, we studied the impacts of capillary effects on the liquid transfer induced by the rupture through numerical simulations. To depict the capillary effects of a porous tip, a time scale ratio, $R_T$, is proposed to compare the competing mechanisms of spontaneous imbibitiona and external drag. In terms of $R_T$, we then develop a theoretical model for estimating the liquid retention ratio considering the geometry, porosity and wettability of tips. The mecahnism presented in this work provides a possible approach to control the liquid transfer with better accuracy in microfluidics or microfabrications.

# Introduction

The dynamics of a liquid bridge formed between two surfaces are referred to as a fluid-solid interaction dominated by capillary effects. Specifically, formation and breakup of liquid bridges can be controlled by kinetic conditions, fluid properties, surface wettability and morphology [1-5]. Although the morphology of adhered surfaces plays an important role on the movement of three-phase contact line [6-7] there are increasing interests on achieving sufficient understanding of their effects on the rupture of liquid bridges [8]. Especially for cases with permeable surfaces, the competing mechanisms of spontaneous imbibition and menisci deformation can raise interesting phenomena [9-11].

Liquid bridges are ubiquitous in nature and well-studied in various industries. In wetted granular materials, e.g., wet sandy soil and humid powders, the capillary force generated by liquid bridges induces the grain-grain cohesion resulting in the increment of shear strength [12], grain agglomerations [13] and self-assembly [14]. With industrial applications, capillary forces generated by the liquid meniscus between a gripper and objects can be used to handle tiny components [15], e.g., watch bearings [16]. This capillary gripping presents significant advantages over the tiny-scale manipulations, such as the indirect contact with the objects, self-centering, high compatibility with various shapes and materials. However, it is required that the remaining liquid on the object be as little as possible [8]. Inversely, for the atomic force microscopy (AFM) technique, the liquid bridge between the probe and object resulting from humid ambience should be eliminated as much as possible to enhance the imaging resolution [17]. In dynamic conditions, liquid bridges can be also formed between the dip and substrate in a variety of transfer printing [18], e.g., off-set printing [19], gravure [20], lithography [21] and so on. The basic technique for printing is to control the liquid volume transferred from one surface to another. In microfluidics, such as used in chemical engineering and biological sciences, there is a general need to accurately manipulate a droplet, and liquid bridges are usually adopted as a tool to merge or transport droplets [22-23]. In these applications, to design a controllable liquid transfer between two surfaces is a basic requirement for various industrial applications.

To date, there have been numerous works investigating mechanisms of rupture-induced liquid transfer between two solid substrates. Specifically, this phenomenon can be divided into two regimes according to the stretching speed [24-25]. One is the dynamic regime where viscous and inertial forces dominate and therefore the liquid bridge can be almost equally separated regardless of other conditions; the other is the quasi-static regime where capillary force dominates so the liquid transfer is determined by the combination of surface wettability. Many existing works mainly focuses on the simple geometries, i.e., plane to plane [26], sphere to sphere [27-28] or sphere to plane [29-30]. A recent work by Tourtit et al. [8] suggests that the surface geometry is another controlling condition for the quasi-static regime, and the relationship between the tip angle of a gripper and its liquid retention has been established.

In many applications, substrates for liquid transfer are often porous. Porous media, e.g., paper, open-cell foams and fabrics, usually work as desirable multiphase flow controllers since their capillary effects can be tuned by changing porous properties, i.e., porosity, pore size, etc [31-32]. Therefore, the controllability of porous media has led to a series of applications, like paper-based chips [33-34], gas diffusion layers in fuel cells [35]. Based on our previous work [36] which concludes that the surface geometry together with the porous structure determines the behavior of a sessile droplet moving on a porous surface, here we further explore the porous structure as grippers for manipulating the dynamics of liquid bridges. The focus of our work is to investigate how porous properties influence the liquid transfer induced by the liquid bridge rupture within a quasi-static regime. This paper is arranged as follows: the numerical scheme is first introduced and validated against an experimental work; then, by combining main controlling factors including stretching speed, surface wettability and porous properties, a scaling parameter, the time ratio $R_T$, is proposed is proposed to compare the two main competing mechanisms of spontaneous imbibitiona and external stretching, and an estimation of liquid retention ratio $L_r$ in terms of $R_T$ is developed based on the simualtion results. Finally, controlling liquid transfer with a better accuracy and wider range by using this development is discussed.

## Numerical Method

To exactly capture the capillary effect within porous zone, we implemented pore-scale resolved models using the phase field numerical scheme. The porous media are realized by a series of homogeneous arrangements of 2D circular obstacles to explicitly model the pore structures. The interface movement and splitting inside and outside the porous zone and its adhesion with solid boundaries are captured based on the phase field method.

## Phase field method for multiphase flow

The phase field method combines Navier-Stokes equation with Cahn-Hilliard diffusion equation [37],

$$\nabla \cdot \boldsymbol{u} = 0 , \tag{1}$$

$$\frac{\partial (\rho \boldsymbol{u})}{\partial t} + \nabla \cdot (\rho \boldsymbol{u}\boldsymbol{u}) - \nabla \cdot (\mu \nabla \boldsymbol{u}) - \nabla \boldsymbol{u} \cdot \nabla \mu = -\nabla p + \boldsymbol{F}_{st} + \boldsymbol{F}_{ext} , \tag{2}$$

$$\frac{\partial \vartheta}{\partial t} + \boldsymbol{u} \cdot \nabla \vartheta = \nabla \cdot \frac{\gamma \lambda}{\varepsilon^2} \nabla \psi , \tag{3}$$

$$\psi = -\nabla \cdot \varepsilon^2 \nabla \vartheta + (\vartheta^2 - 1)\vartheta , \tag{4}$$

where $p$ is the pressure; $\boldsymbol{u}$ is the fluid velocity field; $\mu$ is the fluid viscosity of fluid; $\boldsymbol{F}_{ext}$ is the external body force, e.g., the gravity; $\vartheta$ is the phase variable, which varies in [-1, 1], i.e., $\vartheta = 1$ in the pure gas phase and $\vartheta = -1$ in the pure liquid phase; $\gamma$ is the mobility parameter; $\psi$ is a modified chemical potential that decomposes a fourth-order equation into two second-order equations; $\lambda$ is the mixing energy density; and $\varepsilon$ is a control parameter for the interface thickness that scales with thickness of the interface. The parameters $\lambda$ and $\varepsilon$ are related to surface tension σ through the equation,

$$\sigma = \frac{2\sqrt{2}\lambda}{3\varepsilon^2}, \tag{5}$$

and the surface tension effect can be considered as a body force $\boldsymbol{F}_{st}$, as

$$\boldsymbol{F}_{st} = \frac{\lambda \psi}{\varepsilon^2} \nabla \vartheta. \tag{6}$$

In the phase field model, the interfacial thickness $\varepsilon$ and mobility $\gamma$ are two particularly important parameters. A smaller interfacial thickness $\varepsilon$ requires typically a much finer mesh, thus leading to a great increase in computational cost and causing difficulties in convergence with the phase field method, though it would be close to the solution with sharp-interface assumption. Therefore, the value of $\varepsilon$ should be related to the current mesh size. For any given value of $\varepsilon$, according to the expression of surface tension $\sigma$, the mixing energy density $\lambda$ can be obtained by Eq. (5). The mobility parameter $\gamma$ determines the time scale of the Cahn-Hilliard diffusion, and it thereby governs the diffusion-related time scale for the interface. A suitable value for $\gamma$ is the maximum velocity magnitude occurring in the model and a higher mobility is much helpful to obtain the correct pressure variation crossing the interface. In the following simulations, we use $\varepsilon = 0.1$ mm and mobility $\gamma = 0.1$ m·s/kg.

The wettability condition of solid boundaries, in this numerical scheme, can be applied by two steps: 1) truncate phase diffusion along the normal of solid boundaries; 2) modify the local normal to the interface contacting the solid boundaries, and specifically by implementing

$$\boldsymbol{n} \cdot \nabla \psi = 0 \ ; \tag{7}$$

$$\boldsymbol{n} \cdot \frac{\nabla \vartheta}{|\nabla \vartheta|} = \cos(\theta) \ , \tag{8}$$

where $\theta$ is the contact angle measuring within the liquid phase. The Eq (1)~(4) and its boundary and initial conditons are implemented in the comercial FEA software, Comsol, and additonally the motion of solid boudaries can be realized by the help of moving mesh module.

**Validation**

The recent experimental work by Tourtit et al [8] systematically presented the rupture of a liquid bridge between a solid tip and a plane, and thus enables a clear posibility to validate the numerical scheme. Here, the rupture process of liquid bridge is tracked in a 2D computational domain, and related fluid properties and environmental conditions can be referred in [8]. Though an axisymetric domain seems to be more suitable to model these cornical tips, we did not see the difference from the results obtianed in the current domain and axisymetric approaximations. The former is adopted since later while modelling porous media, the 2D simpilifcation can straightforwardly reflect the porous structure. The liquid rentention ratio $L_r$, i.e., a ratio of liquid volume left on the tip and initial droplet volume when the liquid bridge ruptures, with respect to different tip angles is presented in Figure 1 and compared with the experimental data. As shown in Figure 1, most numerical results are in good agrrement with experimental results except the cases with the tip angle larger than 200° while the experimental data also shows higher variance for these cases. Overall, our simulation results can follow the experiemntally observed behaviours demonstrated in Ref [8], and furthermore the phase field method with moving mesh technique can capture the dynamics of liquid bridges.

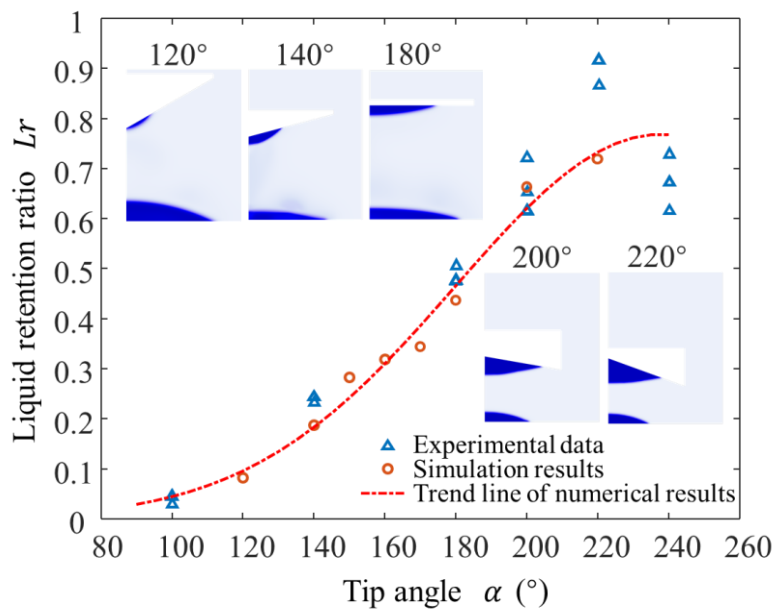

Figure 1. Numerical validation against experimental data for liquid retention ratio $L_r$ with respect to the tip angle $\alpha$.

**Modeling the porous structure and flow conditions**

In this study, we mainly investigate the effect of the porous structure on liquid retention, and therefore fluid properties of the liquid-gas system are fixed in all numerical cases while varying the porous geometry and wettability. A schematic of the present numerical model is shown in Figure 2. Our simulation is on a 2D symmetric domain including a porous zone that is dry initially. The liquid droplet with an initial radius $r_d = 5$ mm is dropped on the solid plane which is moved along the vertical direction. The solid plane moves upwards a certain distance (1 mm) to guarantee the droplet touches the porous tip and the liquid bridge forms between the solid and porous substrates. Then the solid plane moves downwards at a constant speed until the liquid bridge breaks up. A typical water-air system is assumed that liquid viscosity $\mu_l = 1 \times 10^{-2}$ Pa·s and density $\rho_l = 1000$ kg/m³ while gas viscosity $\mu_g = 1.85 \times 10^{-5}$ Pa·s and $\rho_g = 1.2$ kg/m³; the surface tension $\sigma$ between these two fluids equals 72.9 mN/m; to guarantee the capillary effect controls the whole process instead of the viscosity, the moving speed $u$ of the solid plane is kept lower than 4 mm/s so that the capillary number $Ca = \frac{\mu_l u}{\sigma} < 5.6 \times 10^{-5}$.

As for the geometry of porous tips, microscopically, the circular obstacles inside the porous tips are arranged on regular hexagonal grids, which can be considered as the simplest representation of homogeneous porous media. Here, we fix the centre-to-centre distance $l_c = 1$ mm between two neighbour obstacles and change the obstacle size to adjust the porosity $\phi$ (0.42 and 0.49) and correspondingly throat size $d$ (0.2 mm and 0.25 mm) of porous zones; macroscopically, the shape of porous zone varies from wedge to plate and is controlled by the tip angle $\alpha$ ranging from 60° to 180°, correspondingly. Additionally, the contact angle of the porous zone $\theta$ covers a range of 30°, 45° and 60° while the contact angle for the solid plane is fixed as 30°. It is expected towards the hydrophilic conditions, the porous tips will behave as solid ones since the liquid penetration will be impeded.

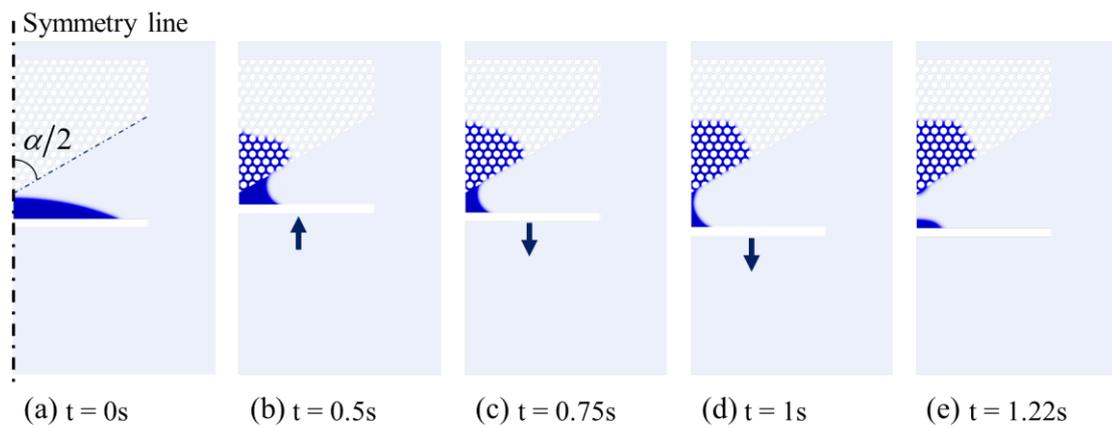

Figure 2. Schematic of the simulation model for porous tips and a typical evolution of liquid bridge with the loading time with $\alpha = 120°$ and $u = 4$ mm/s.

## Results and Discussion
### Simulation results

By employing the proposed numerical model, the dynamics of a liquid bridge are simulated with different geometries and conditions. As shown in Figure 2, after touching the porous tip, the droplet infiltrates the porous zone driven by the capillary pressure and the apparent contact line outside the tip move up along the oblique side with the inside liquid front forming in a liquid bridge; with the solid plane being dragged far away from the porous tip, the contact line slips on the porous surface and tends to pinch while the liquid continues to imbibe; once the liquid bridge breaks up, part of the outside liquid remain on the tip while the other part is still attached on the plane. The ration of the latter volume to the initial volume is defined as the liquid retention ratio $L_r$.

The simulation results, as shown in Figure 3, present that the liquid retention ratio $L_r$, covering a wide range from less than 0.4 to more than 0.9, is firstly controlled by the porous structure (porosity $\phi$). Compared Figure 3(a) to (b), the contour lines shift to the top-left corner, i.e., $L_r$ increases slightly for $\phi$ changes from 0.42 to 0.49 since the characteristic imbibition time determined by the combination of the permeability of porous zone $k$ and capillary pressure $P_c$ [38], goes down correspondingly. Meanwhile, these two figures show similar trends of $L_r$ for the combination of contact angle $\theta$ and tip angle $\alpha$, i.e., the decrease of $\theta$ and increase of $\alpha$ result in more liquid retained on the tip. Specifically, for given porosity, the smaller contact angle, i.e., the porous zone is more hydrophilic to liquid, leads to a higher capillary pressure, or a lower characteristic imbibition time, that drives the liquid infiltrates the porous tip. Together with porosity $\phi$, contact angle $\theta$ determines the capillary effects on the dynamics of liquid bridge. For the tip angle, macroscopically it controls the slide of contact line, as discussed in Ref [8], i.e., the smaller $\alpha$ is, the easier for the contact line to move along the tip surface so that less volume of liquid would be retained on the tip at the moment of the rupture of a liquid bridge; microscopically, $\alpha$ also determines the diffusion area which is particularly discussed in the following section. Figure 4 shows the variation of capillary forces applied on the solid object as a function of the stretching distance and speed until ruptures of the liquid bridge when the capillary force jumps to zero. With the tip angle increasing, the magnitude of capillary forces increases while the rupture distance decreases. Also, the reduction range of capillary force from the initial state to rupture increases. Especially for $\alpha = 60°$, the capillary force almost keeps a constant during the stretching since the contact lines adhered on the porous and solid surface slide at a similar pace so that the curvature of the gas-liquid interface does not change. Comparing Figure 4 (a) and (b), slowing the stretching speed results in the reduction of the rupture distance since imbibition the as a competing factor of stretching dominates the liquid bridge rupture, specifically more liquid infiltrates the porous zones during a relatively slower stretching process. This competition can be described by two characteristic time scale, as discussed in the following section.

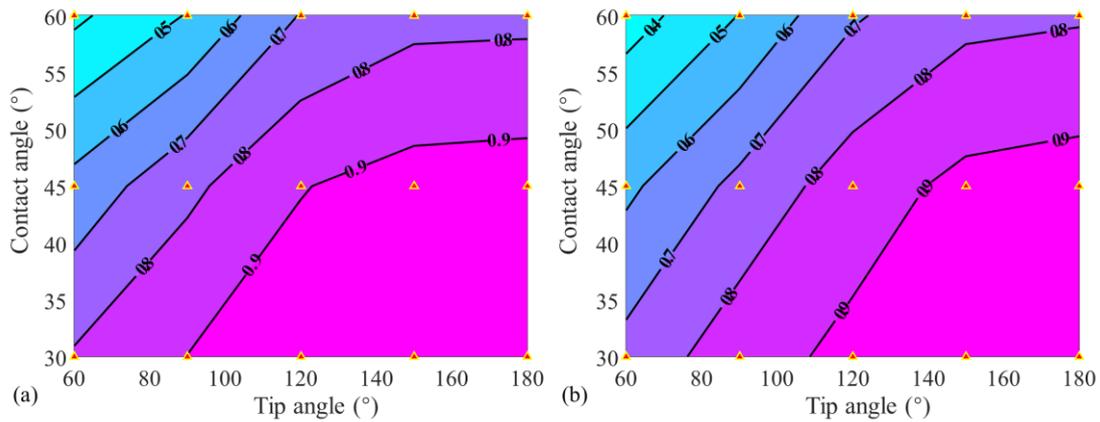

Figure 3. Smoothed contour plots of liquid retention ratio with changing contact angle $\theta$ and tip angle $\alpha$ for different porosities $\phi$, (a) 0.42 and (b) 0.49. The markers (▲) indicate the simulation cases.

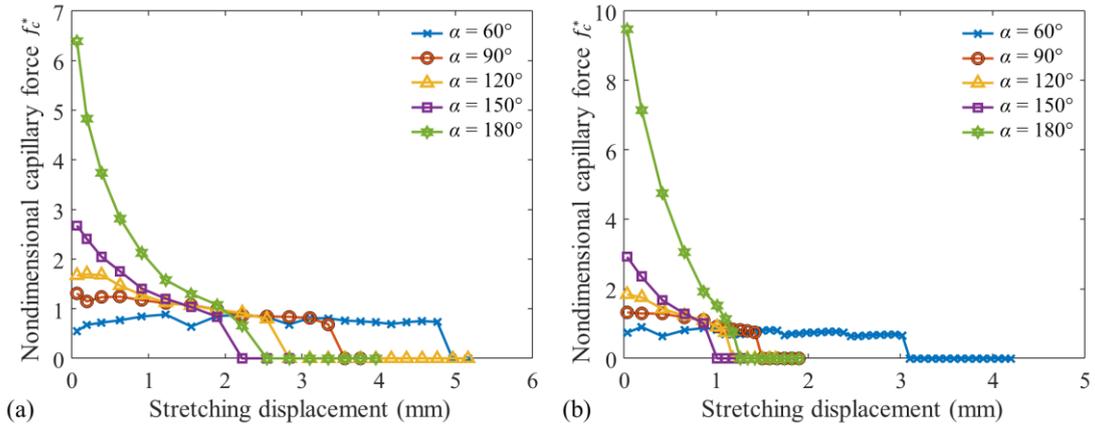

Figure 4. Nondimensional capillary force $f_c^*$ vs. stretching displacement with changing tip angle $\alpha$ for different porosities stretching speed $u$, (a) 4 mm/s and (b) 0.8 mm/s, with porosity porosities $\phi = 0.42$ and contact angle $\theta = 45°$. Here, $f_c^*$ is a ratio of per-unit-length capillary force of the moving plate and surface tension.

## Dimensionless analysis

### a) Capillary-driven imbibition

Assuming the contact area between the droplet and the porous tip as a point liquid source and the liquid front as a circular segmented centred on the point source, as shown in Figure 5, the infiltration area is

$$S_{imb} = \frac{1}{2}\phi H^2 \alpha, \tag{9}$$

and according to the Lucas-Washburn equation [38-39], within a homogeneous porous media, the position evolution of liquid front $H$ with time $t$ can be estimated as

$$H^2 = \frac{2kP_c^*}{\phi \mu_l} t, \tag{10}$$

where $k$ and $P_c^*$ is the permeability and characteristic capillary pressure of the porous tip. For the hexagonal arrangement of obstacles, as studied here, the permeability is a function of $\phi$ and $d$, and can be estimated as $k = d \cdot \frac{\phi^{0.76}}{(1-\phi)^{0.24}}$ [32]; microscopically, the capillary pressure $P_c$ is related to the contact line movement, i.e., filling angle $\varphi$, and the representative value of capillary pressure $P_c^*$ can be estimated as a mean value of $P_c$, i.e.,

$$P_c^* = \frac{\int_{\alpha_0}^{\pi/2} P_c(\varphi)d\varphi}{\pi}. \tag{11}$$

More details about Eq. (11) can be referred to [32].

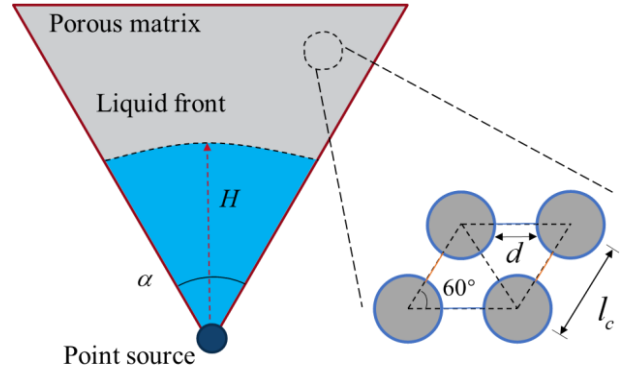

Figure 5. Schematic of a liquid-infiltrated porous tip with tip angle $\alpha$

**b) Time scales**

The stretch velocity and capillary imbibition are two dominating factors on the liquid retention of porous tip, and they can be related to respective characteristic time scale, i.e., macroscopic time length $T_M$ and microscopic one $T_m$. A time ratio $R_T$ as dimensionless number is proposed here to link geometrical and flow conditions together. Specifically, $T_M$ is the characteristic time during which the bottom solid substrate is dragged through a distance equalling to the equivalent droplet size,

$$T_M = \frac{r_d}{u}, \tag{12}$$

where $r_d$ is the initial size of a droplet; $T_m$ is the time for the whole droplet imbibes by capillary suction of the porous tip, and thus setting the imbibition area $S_{imb}$ equal to the droplet area, i.e., $\frac{1}{2}\pi r_d^2$. Combining with Eq. (10), $T_m$ can be solved as

$$T_m = \frac{\pi r_d^2}{\alpha} \frac{\mu_l}{2kP_c^*}. \tag{13}$$

Then, the original time scale ratio $R_T^0$ is expressed as

$$R_T^0 = \frac{T_M}{T_m} = \frac{2\alpha}{\pi r_d u} \frac{kP_c^*}{\mu_l}. \tag{14}$$

Considering the deviation resulting from the assumption of the point source, a correction term related to the droplet size is proposed here,

$$R_T = R_T^0 \left(\frac{r_d}{d}\right)^n, \tag{15}$$

where $n$ is a fitting index. Based on the Eq. (15), a larger $R_T$ suggests that the external drag becomes slower or it takes less time for liquid to infiltrate into the porous tip, so the proposed $R_T$ should be positively correlated with liquid retention ratio, $L_r$.

c) **Liquid bridge rupture**

As suggested in reference [8], with the contact line movement on the solid surface, the liquid bridge finally break up and part of liquid will be left on the tip, and the volume of the leftover is a function of tip geometry and surface property. However, for a porous tip, the imbibition inside the porous zone and outside stretching, as two competing factors, together impact the rupture of capillary bridges and liquid retention furtherly, as shown in Figures 3 and 4. Similarly, the outside residual liquid induced by the capillary bridge breakup, contributing to the liquid retention ratio, is determined by tip angle $\alpha$ and contact angle $\theta$, and can be denoted as $R_S$.

As analysed above, the liquid retention ratio for a porous tip can be estimated as

$$L_r = R_S + (1 - R_S)f(R_T), \quad (16)$$

which includes two parts, i.e., the first term is the contribution of liquid bridge breakup and the second term is the additional retention induced by interior capillary effects. For the capillary part, specifically, when $R_T = 0$, suggesting that the capillary effect can be ignored and the liquid retention is identical to the one of smooth solid tip, $f(0)$, therefore, should be zero; when $R_T \to +\infty$, i.e., the relative movement of porous tip tend to be motionless ($u \to 0$), the capillary-induced increment reach the upper limit and $f(+\infty)$ should be one. Thus, $f(R_T)$ can be formulated as

$$f(R_T) = \frac{e^{R_T/R_T^*} - 1}{e^{R_T/R_T^*} + 1}, \quad (17)$$

where $R_T \in [0, +\infty)$, and correspondingly $f(R_T) \in [0,1)$. After the nonlinear regression analysis based on the all obtained simulation data, containing 45 cases covering variations of tip geometry, wettability, and porous structures, the parameters $R_T^*$ and $n$ in the proposed correlation are fitted as 875.1 and 1.5, respectively, with $R^2 = 0.89$.

In summary, the proposed expression regarding liquid retention ratio in Eq. (16) combines two competing factors, i.e., imbibition inside the porous zone and contact line movement outside the porous zone. With the help of Eq. (16) we can predict the liquid retention ratio for given geometry and flow conditions, and furthermore, for a given porous tip, we can also adjust the stretching speed to satisfy various liquid retention requirements. Inversely, the tip geometry, including tip angle and inner porous structure, can be designed based on the proposed relationship to meet different application situations. A typical scenario can be implementation of multiple porous tips, with different design parameters (e.g., porosity and tip angle), in a system simultaneously to selectively manipulate droplets.

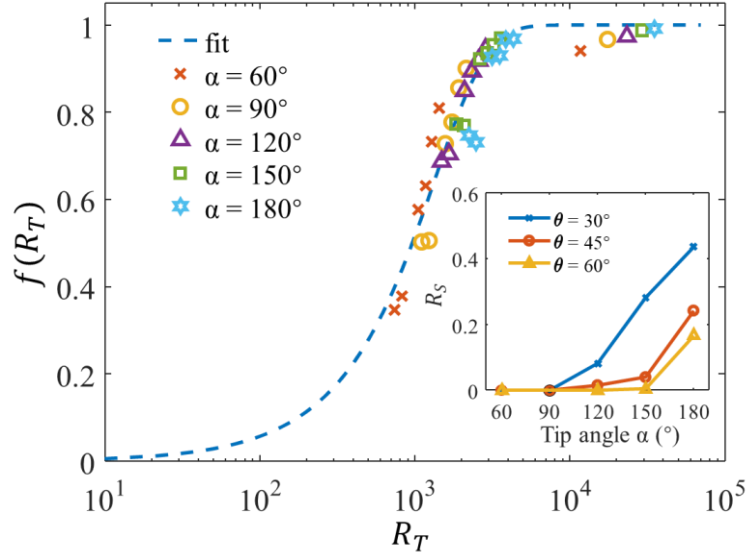

Figure 6. Simulation results and nonlinear regression curve of weight function $f(R_T)$. The insert presents the change of the contribution of liquid bridge breakup to the liquid retention ratio $R_S$ with the tip angle $\alpha$ and contact angle $\theta$.

## Conclusion

In this work, we studied the dynamics of a liquid bridge between a porous tip and a moving solid plane, and the impacts of capillary effects on the liquid transfer was investigated through a series of numerical simulations. Specifically, there are generally two volumes contributing to the liquid retention during the liquid bridge rupture. One is the imbibition into the porous tip, i.e., the initial droplet, as a liquid source, infiltrates the porous zone during stretching; the other is the contact line movement outside the porous tip, i.e., a part of outside liquid is retained on the tip once the rupture occurs. We proposed a scaling parameter $R_T$, a time scale ratio, to quantify the relative contribution from capillary effects and then an estimation of liquid retention ratio $L_r$ considering these two contributions is developed based on the simulation results. Notably, our estimation of $L_r$ can keep consistent with the conclusion for solid tips (i.e., $R_T = 0$) reported in Ref [8], whilst the proposed empirical equation can capture the additional contribution from the porous substrates. This study on rupture of liquid bridges suggests another potential way to accurately control liquid transfer and manipulate droplets required by microfluidics and microfabrication.